\documentclass[journal]{vgtc}                

\ifpdf
  \pdfoutput=1\relax                   
  \usepackage{graphicx}                
  \DeclareGraphicsExtensions{.pdf,.png,.jpg,.jpeg} 
\else
  \usepackage{graphicx}                
  \DeclareGraphicsExtensions{.eps}     
\fi%

\graphicspath{{figures/}{pictures/}{images/}{./}} 

\usepackage{microtype}                 
\PassOptionsToPackage{warn}{textcomp}  
\usepackage{textcomp}                  
\usepackage{mathptmx}                  
\usepackage{times}                     
\usepackage{cite}                      
\usepackage{tabu}                      
\usepackage{booktabs}                  
\usepackage{xcolor}

\onlineid{0}

\vgtccategory{Research}
\vgtcpapertype{evaluation}

\title{
What User Behaviors Make the Differences \\ During the Process of Visual Analytics?
}
\author{
Zekun Wu \thanks{e-mail: zwu13@uncc.edu}, 
Shahin Doroudian \thanks{e-mail: sdoroudi@uncc.edu}, 
and Aidong Lu \thanks{e-mail: aidong.lu@uncc.edu}
}

\authorfooter{
}

\abstract{
The understanding of visual analytics process can benefit visualization researchers from multiple aspects, including improving visual designs and developing advanced interaction functions.
However, the log files of user behaviors are still hard to analyze due to the complexity of sensemaking and our lack of knowledge on the related user behaviors.
This work presents a study on a comprehensive data collection of user behaviors, and our analysis approach with time-series classification methods.
We have chosen a classical visualization application, Covid-19 data analysis, with common analysis tasks covering geo-spatial, time-series and multi-attributes.
Our user study collects user behaviors on a diverse set of visualization tasks with two comparable systems, desktop and immersive visualizations.
We summarize the classification results with three time-series machine learning algorithms at two scales, and explore the influences of behavior features.
Our results reveal that user behaviors can be distinguished during the process of visual analytics and there is a potentially strong association between the physical behaviors of users and the visualization tasks they perform.
We also demonstrate the usage of our models by interpreting open sessions of visual analytics, which provides an automatic way to study sensemaking without tedious manual annotations.
} 


\CCScatlist{
  \CCScatTwelve{Human-centered computing}{Visu\-al\-iza\-tion}{Visu\-al\-iza\-tion techniques}{};
  \CCScatTwelve{Human-centered computing}{Visu\-al\-iza\-tion}{Visualization design and evaluation methods}{}
}

\teaser{
\centering
\includegraphics[width=3.2in]{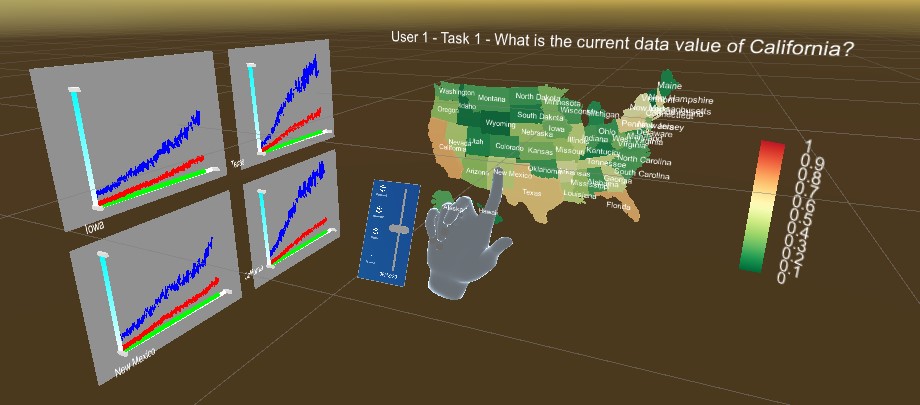}
\includegraphics[width=3.2in]{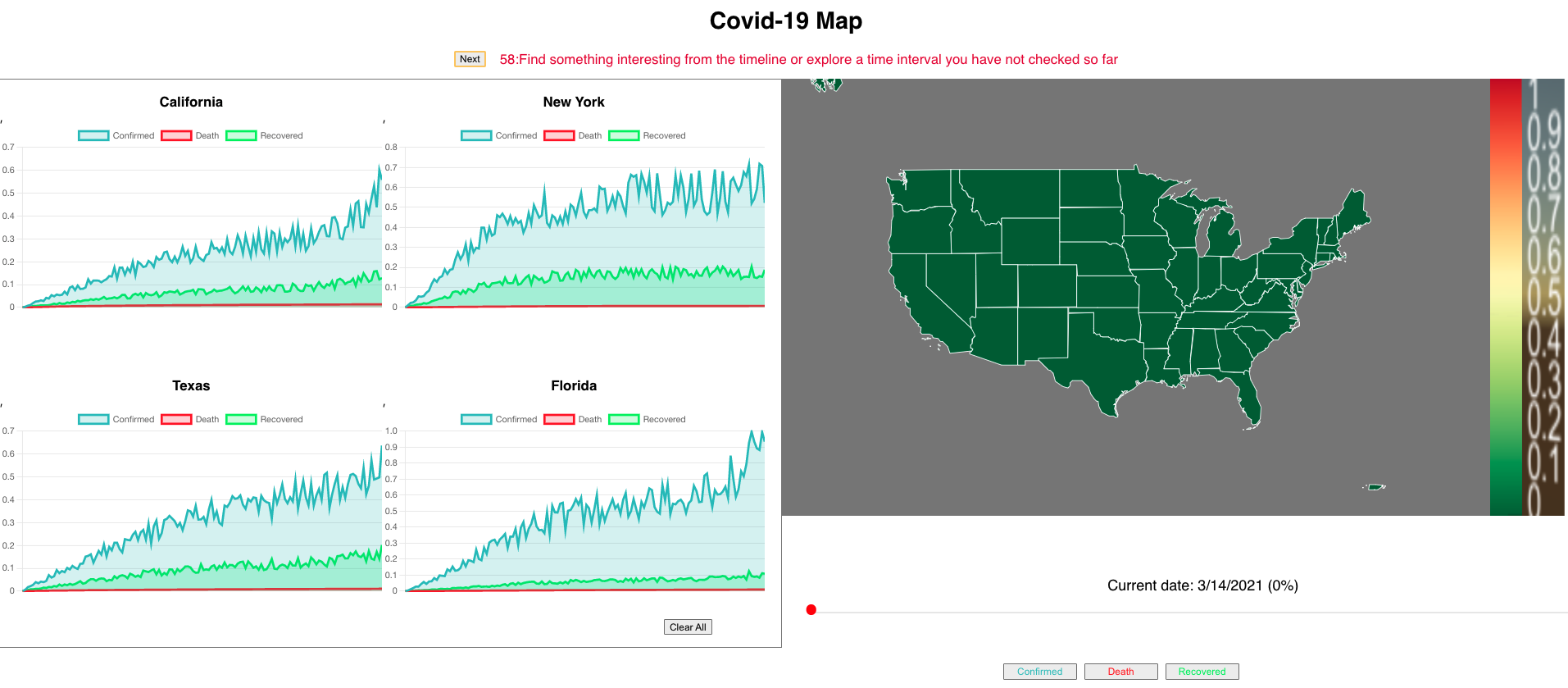}
\caption{This work examines user behaviors during the process of visual analytics with both desktop and immersive visualizations. The two comparable systems both contain three visualization panels (choropleth,  time charts and interaction panel) and the same set of interaction functions implemented with different input methods of the two platforms.}
\label{fig:systems}
}

\vgtcinsertpkg

\begin{document}


\firstsection{Introduction}

\maketitle


Previous work has revealed the complexity of user behaviors during the process of visual analytics~\cite{7192714, Xu2020}.
Up to now, the sensemaking process in visualization still remains as a challenging problem~\cite {Pirolli2005,Crooks1988, Andrews:2010:STL:1753326.1753336}.
The problem is hard to study, since it is an internal process of mind that we cannot retrieve directly. 
Instead, we can only rely on external user behaviors to reveal the internal process.
Previous works in visualization have focused on analyzing interaction provenance, and many rely on manual or semi-automatic approaches to create sensemaking models or annotate the visualization processes~\cite{7192714, Xu2020}.
However, the small collection of user behaviors and the lack of automatic methods limit the results on the process of visual analytics.

There are also interesting findings that show the sensemaking process varies depending on external factors, such as the environments with different display sizes and resolutions~\cite{10.1145/2702123.2702406}, visualization systems with latency~\cite{https://doi.org/10.1111/cgf.13678}, or experiences of users~\cite{7539616}.
All the factors contribute to our understanding of the visualization process, and it is beneficial to enrich this collection.
To the best of our knowledge, previous work focused on the attributes of interaction provenance, and the factors of physical behaviors of users (such as body movements) have not been fully studied yet.

The focus of this work is to explore two fundamental research questions on understanding user behaviors during the general process of visual analytics.
\begin{itemize}
\item If external user behaviors from different tasks of visual analytics can be distinguished without requiring the annotations from visualization systems or users?
\item Are the results different between desktop and immersive visualizations based on different collections of behavior features?
\end{itemize}

To answer these questions, this work presents a new data-driven approach to study the process of visual analytics by combining the latest methods of immersive analytics and machine learning.
We choose a classical visualization problem, with geospatial, temporal and multi-attribute visualizations, to study the general processes of visual analytics.
We have designed a set of basic visualization tasks based on information task taxonomies, which serve as the automatic annotations and candidates for distinguishing general user behaviors.

Our study collects a rich set of user behaviors at high frequencies, including attributes from general interaction provenance and eye tracking data and physical movements of users that are used to require specialized devices.
To study the differences between desktop and immersive visualizations, we have developed two comparable systems with the same set of visualizations and interaction functions.
Accommodating different interaction channels from the two environments, we organize our data collection as three groups of behavior features.
The interaction and selection groups are shared by the desktop and immersive visualizations, and the immersive group is unique to the latter.

We have surveyed machine learning methods for time-series data, and chosen three classification models, kNN, CNN and ROCKET.
For efficiency of model training, we perform data transformation and cleaning to preserve temporal and statistical features of behavior patterns.
We summarize the classification results for the two environments at two scales, the low-level task category and the high level visual space.
Our results reveal that the user behaviors during the process of visual analytics can be distinguished (around $90.0\%$ accuracy score for the immersive environment).
We also identify the influences of individual behavior features on the classification models and demonstrate that physical behaviors of users are connected to their sensemaking process.

To demonstrate the results of our classification models, we show how our models can be used to interpret the visual analysis process during open tasks.
By matching the recorded voices from the study, we show that our results enable researchers to understand the sensemaking process in an automatic way, which is much more efficient compared to the manual labeling approaches and allows us to collect large training sets for diverse visualization tasks from multiple participants.

This paper is organized as follows:
We first present the related works on the process of interactive exploration and sensemaking, and analysis of interaction provenance in Section 2.
Section 3 provides an overview of our study, including the study design and selection of visualization tasks.
Section 4 presents our two comparable visualization systems, and Section 5 describes the details of the user study.
We present our data transformation and classification models in Section 6 and data analysis and results in Section 7.
Section 8 discusses our work and presents the future work.

\section{Related Work}

We review related work on the interactive exploration process of visual analytics, and collection/analysis methods of interaction provenance.

\subsection{Interactive Exploration and Visual Analytics}

The processes of interactive exploration and sensemaking in visualization and visual analytics have been studied extensively.
Previous studies have not only shown approaches to study such processes, but also demonstrated several important usages to understand users and create advanced visualization functions.
There are also evidence on the common features of analyses processes across participants, as well as potential structural differences between behavior graphs for open-ended and focused exploration tasks~\cite{https://doi.org/10.1111/cgf.13678}.

One thread of the main efforts is to create conceptual or computational models to capture the interactive processes, including graphs, Markov chain, action tiers and cognition models~\cite{doi:10.1057/ivs.2008.28}.
Among which, graphs have been used often to represent various transitions of system and users during the processes, such as the transformation among visualization sessions~\cite{1183791,4069243}.
The Markov chain process was used to model visual exploration to show the transitions among mental, interaction, and computational states
~\cite{doi:10.1177/1473871616638546}.
Action tiers (tasks, sub-tasks, actions and events) were used to bridge the gap between semantic interactions and system events~\cite{4677365}.
Activity Theory was also used as an organizing framework for modelling human activities and constructing activity typologies for visual analytics~\cite{8019880}.
A mental model constructed from top-down reasoning process was also used to understand the human interactions~\cite{5613437}.

What makes the interactive exploration processes hard to model comes from multiple factors, and one is the multiple-levels feature of such processes.
Through observing how a visualization task was performed and the related task inputs and outputs, multiple levels of tasks were used to abstract visualization tasks with how, why and what questions~\cite{6634168}.
Similarly, multiple tiers~\cite{4677365} were observed from analytic behaviors.
The exploration processes can also be affected by other factors such as display size and resolution~\cite{10.1145/2702123.2702406}.

The interactive exploration process has also been studied in immersive environments, where virtual or augmented realities are often used to explore the cognitive activities during visual analytics processes~\cite{Lee:2012:BMK:2720013.2720425, 7390081, 8017617, Cordeil:2017:IIA:3126594.3126613, 10.1111:cgf.13431,8440844}.
Different from desktop systems which use mouse and keyboard, interaction functions in immersive systems are often achieved through multi-sensory channels including gaze, voice, and gestures.
During the use of immersive systems, 
the users are more likely to mix their interactions with the visualization systems with their physical movements.
Since the field of immersive visualization is still young, summative evaluations of the advantages or disadvantages compared to desktop systems have not been achieved yet.
Our understanding of the interactive exploration process between the two types of systems may help us to design effective immersive visualization and choose the suitable platform. 


The models of the interaction processes can be used to understand user behaviors, such as external anchoring, information foraging, cognitive offloading~\cite{5613437}, enumerate several exploration strategies~\cite{8281629, 9382916}, prediction of skill levels~\cite{10.1145/2557500.2557524} and evaluation~\cite{7192662,LIVVIL}.
In addition, the results can be used for automated reporting, analysis, or visualization~\cite{1183791} and generate guidance~\cite{Ceneda2019}.
Vistrails~\cite{1532788} demonstrated that new visualization tools could be built by combining the records from the visualization process.   
When users were aware of their interaction history, it could impact users' exploration and insights~\cite{7539616}.
The multi-level task structures were used to create summaries of charts for effective collaboration and multi-user analyses~\cite{Xu2018ChartCE}.

The main advantage of these methods is `open box', which reveals intricate processes with models that we can further annotate, visualize or interact with for additional tasks. 
Different from these approaches, our work focuses on studying factors of different behavior attributes that make major differences during the interactive exploration processes.

\subsection{Collection and Analysis of Interaction Provenance}

Interaction provenance is a major topic to study the visual analytics process, and can benefit visualization from multiple aspects~\cite{7192714, Xu2020}.
We summarize the related work from the aspects of visualization tasks focused in the studies, two main threads of efforts -- manual and automatic approaches~\cite{4677365}, and data collection. 

The tasks focused in the studies are often the basic visualization and visual analytics questions for common applications.
Both open and close-ended tasks have been studied.
For example, a study focusing on how people discovering the functionality of an interactive visualization that was designed for the general public~\cite{8281629},
instructed the participants to freely explore the views and to talk aloud during the process.
Two types of tasks, ``retrieve value" and ``compute derived value" involved comparing individuals against a group average with bar charts, were chosen as primitive data analysis tasks~\cite{10.1145/3025171.3025187}.
Five basic type of tasks (retrieve value, filter, compute derived
value, find extremum, and sort) and single/double options for different task complexities were considered~\cite{10.1145/2449396.2449439}.

Manual/semi-automatic labeling or annotation approaches were common during the past to create interaction provenance.
For example, a model to demonstrate the usage of provenance information by recording the history of user explorations in visualization and allowing users to annotate their explorations~\cite{Groth:2006}.
Click2Annotate was designed to use a semi-automatic annotation method to capture low-level analytics tasks such as clusters and outliers~\cite{5652885}.
Manual methods were also used to collect explanation along the user interaction in crowdsourced data~\cite{6634191}.
Manual coding was used to annotate reasoning process, which was later feed to machine learning methods to classify user interactions~\cite{doi:10.1177/1555343416672782}.
They showed a significantly improved classification result compared to all manual approaches.
A visual design method was used to build models to predict focus of users with their collected gaze and click data~\cite{7192728}.
Semi-automatic methods improved the data collection process significantly~\cite{loom}.
Interactive visualization approaches have also been developed to identify and annotate patterns for provenance~\cite{7539341, 9472937}.

Automatic statistical and machine learning methods have been shown to out-perform previous methods~\cite{10.1145/2449396.2449439}.
Bayesian classification is used to track a viewer's interests during exploration process~\cite{6133283}.
The K-Reversible algorithm was used to determine the common patterns among a grammar-based model collected from interaction provenance~\cite{7534759}.
Among the provenance data, eye tracking data was used to improve the prediction of users' skill acquisition state with additional features of pupil dilation and head distance~\cite{10.1145/3025171.3025187}.
Eye tracking data have also been used with machine learning classifiers (Decision Trees, Support Vector Machines (SVMs), Neural Networks, and Logistic Regression) for prediction of skill acquisition (among which Random Forest algorithm achieved the best results)~\cite{10.1145/3025171.3025187}, 
infer visualization tasks and user cognitive abilities~\cite{10.1145/2449396.2449439},
prediction of skill levels~\cite{10.1145/2557500.2557524}, study task properties, user performance, and user cognitive traits~\cite{10.1145/2633043},
and real-time prediction of user cognitive abilities in visualization tasks~\cite{10.1145/3301400}.
The Multivariate Long Short Term Memory fully convolutional network was shown to predict visual search task success from eye gaze data~\cite{10.1145/3446638}.

The interaction provenance includes systems logs, audio/video recordings, eye tracking data~\cite{8281629}, and VR/AR data~\cite{9382916}.
Also for think-aloud studies, voice transcripts could be collected from recordings and used to train ML models to detect problem encounters~\cite{8807301}.
The main difference of our work from previous studies is that we collect a comprehensive set of interaction provenance at high-frequency throughout the exploration process, and identify major factors for classification of user behaviors in visual analytics processes.

\begin{table*}
\centering
Tasks Categories Organized by the Visualization Spaces\\
\begin{tabular}{ |p{1.5cm}||p{7.5cm}|p{1.5cm}|p{1.5cm}|p{1.3cm}|p{1.3cm}| }
 \hline
 Vis Space & Task Description & Category No. & Choropleth & Time Chart & Interaction Panel\\
 \hline
 \hline
 Spatial & Tasks only related to the spatial visualization  & (0)0 - (0)9 & * &  &\\
\hline
 Temporal & Tasks only related to the temporal visualization & (1)0 - (1)9 &  & * &\\
\hline
 Interaction & Tasks only related to the system interactions & (3)0 - (3)4 &  &  & *\\
 \hline
 Combined & Tasks involve all the spatial/temporal visualizations and interactions
 & (2)1 - (2)9 & * & * & *\\
 \hline
\end{tabular}
\caption{As $*$ marks, the spatial tasks are performed mainly from the choropleth, temporal tasks from time charts, interaction tasks from interaction panel, and combined tasks from all panels.
The interaction tasks include functions necessary for selecting any data items in the spatial and temporal domains, and the other three spaces all contain 10 basic visualization task types listed in Table~\ref{table:task2} (except task (4)0 ``retrieve value" from a combined space is removed for being similar to (1)0 from spatial space or (2)0 from temporal space). 
The category numbers indicate the (space number) and task type number, and a total of 34 tasks are used in our study.
By covering all individual and combined spaces, we ensure to select a diverse set of visualization tasks.
}
\label{table:task}
\end{table*}

\begin{table}
\centering
Visualization Task Types\\
\begin{tabular}{ |p{1cm}||p{2cm}|p{4.5cm}| }
 \hline
 Type No. & Type Name & Task Description\\
 \hline
 \hline
 0 & Retrieve Value & retrieve a value from visualization\\
\hline
 1 & Compute Derived Value & compute an aggregate numeric representation of a data set\\
\hline
 2 & Find Extremum & find maximum or minimum from a data set\\
\hline
 3 & Sort & rank a data set according to some ordinal metric\\
 \hline
 4 & Determine Range & find the span of values within a data set\\
 \hline
 5 & Characterize Distribution & characterize the distribution of an attribute’s values over the set\\
 \hline
 6 & Find Anomalies & identify any anomalies within data\\
 \hline
 7 & Cluster & find clusters of similar data values\\
 \hline
 8 & Correlate & determine useful relationships between the values of two attributes\\
 \hline
 9 & Exploration & find some new data features\\
 \hline
\end{tabular}
\caption{Visualization Task Types selected from Taxonomy~\cite{1532136} (all close-ended) and an open-ended exploration type.}
\label{table:task2}
\end{table}

\section{Study Design}

\subsection{Challenges and Design of Training Data Collection}

The first challenge of exploring ML models is the collection of training data.
Currently there are no public datasets available to use, and we need to figure out a good mechanism to collect data with quality.
It is ideal to design an automatic training method that does not require any types of manual/semi-automatic labeling, since 1) the training of ML models often requires a good size of data samples and manual labeling would be unpractical and 2) the quality of manual labeling is highly depended on the interpretations of participants.
Therefore, we adopt a supervised training method that collects various user behaviors during a set of visual analytics processes.

Second, what labels are suitable for the training data? 
Since our goal is to study the differences of user behaviors in visualization, we need labels that can serve as a language to describe various visual analytics processes.
As the previous work on interactive visualization processes often ask participants or researchers to annotate user behaviors with the visualization tasks~\cite{7192714, Xu2020}, we search for a suitable collection of visualization tasks that
can offer a good coverage of diverse activities.
Also, we cannot use one study to cover the entire visualization field, and need to select a good visualization application with a set of interesting visual analytics problems.
Specifically, our questions are: 
\begin{itemize}
\vspace{-2mm}
\item Which visualization application is suitable for the study? 
\vspace{-2mm}
\item What visualization tasks are suitable for labeling user behaviors? 
\end{itemize}
\vspace{-2mm}
The following describes the rationales behind our choices on the visualization application and tasks for the training labels.

\subsection{Dataset and Application} 
We choose a US Covid-19 dataset~\cite{coviddata}, which records three attributes -- total confirmed cases, total recovered cases and total demises, for each state, each day for around 2 years (the time range of January 2020 to January 2022).
We have also considered the world wide Covid-19 dataset, but choose the US version for similar sizes of states to reduce the behavior differences from the interactive selection tasks.

The dataset is equipped with the features of common geo-spatial datasets -- multiple data attributes are distributed across the geo-space and time-dimension simultaneously, making geo-spatial analysis tasks challenging to perform effectively.
These features of geo-spatial analysis are very suitable for immersive visualization, which often overlays geo-spatial and temporal dimension in 3D physical environment naturally.
In addition, immersive visualization can be designed to layout all the visualization and interaction panels around the user, improving the distributed cognition and involving physical behaviors of users that could be captured with sensors on immersive devices~\cite{8820171,8533893}. 
It enables us to record various user behaviors related to the sensemaking process and later create efficient model to characterize the behavior features.

Since our study contains multiple tasks, we need to avoid the repetition of the same dataset for different tasks.
Also, we want to maintain the consistent style of the data and the common sense of Covid-19 records.
Therefore, we divide the dataset into several segments with an equal duration.
Each dataset is normalized in the whole duration to keep consistent minimum and maximum values.
We also increase the variety of the data patterns by adding random numbers (up to 20\%) to the data and further adjusting the scales of each state within the minimum and maximum value range of the segment.

\begin{figure*}[h]
\centering
 \includegraphics[width=3.0in]{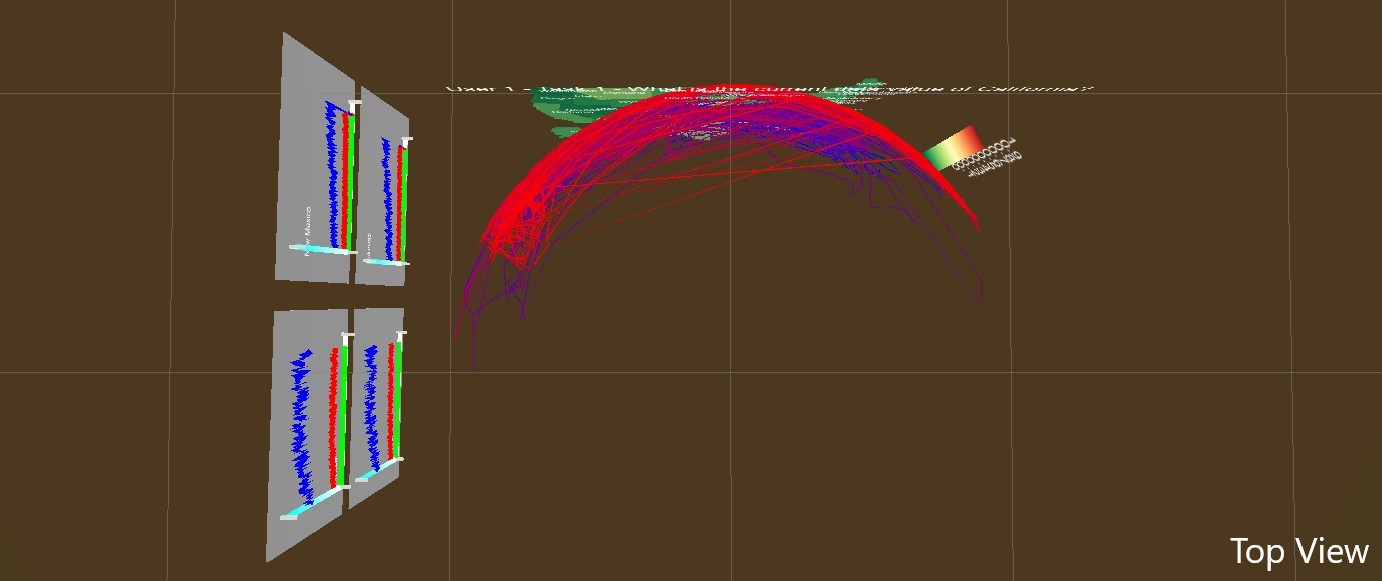}
 \includegraphics[width=3.0in]{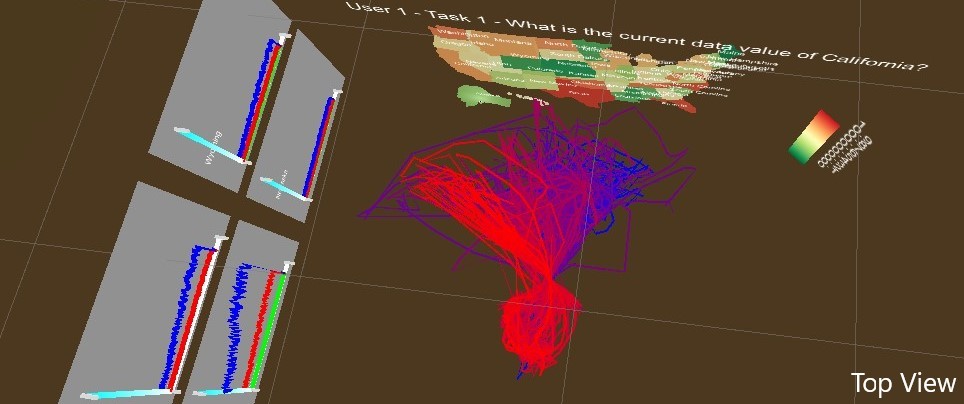}\\
 (a) \ \ \ \ \ \ \ \ \ \ \ \ \ \ \ \ \ \ \ \ \ \ \ \ \ \ \ \ \ \ \ \ \ \ \ \ \ \ \ \ \ \ \ \ \ \ \ \ \ \ \ \ \ \ \ \ \ \ \ \ \ \ \ \ \ \ \ \ \ \ \ \ \ \ \ \ \ \ \ \ \ \ \ \ \ \ \ \ \ \ (b)\\
  \includegraphics[width=3.0in]{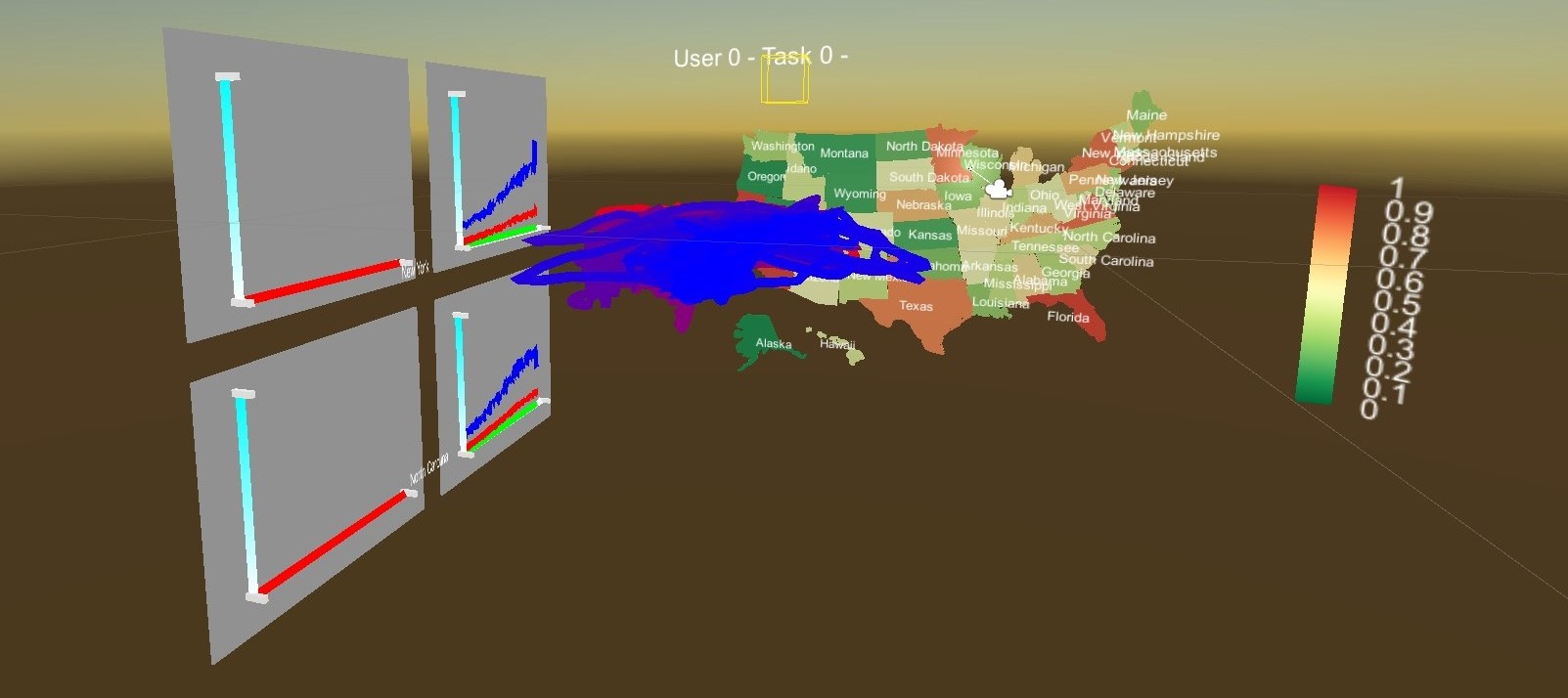}
  \includegraphics[width=3.0in]{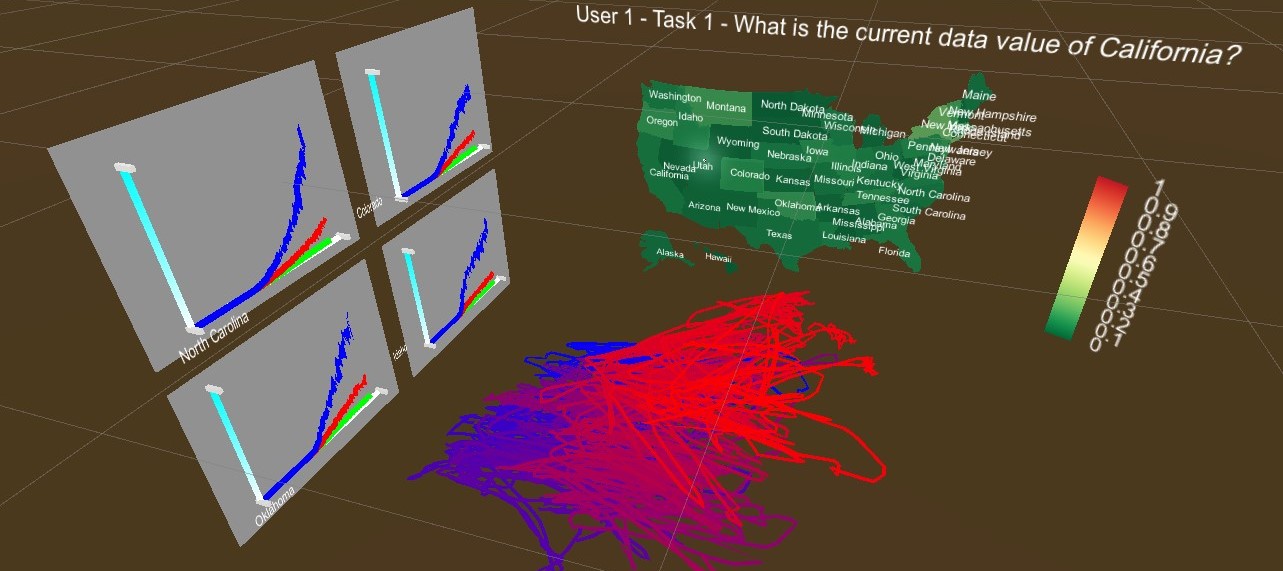}\\
   (c) \ \ \ \ \ \ \ \ \ \ \ \ \ \ \ \ \ \ \ \ \ \ \ \ \ \ \ \ \ \ \ \ \ \ \ \ \ \ \ \ \ \ \ \ \ \ \ \ \ \ \ \ \ \ \ \ \ \ \ \ \ \ \ \ \ \ \ \ \ \ \ \ \ \ \ \ \ \ \ \ \ \ \ \ \ \ \ \ \ \ (d)\\
\caption{We illustrate the behaviors of users when they were performing visual analytic tasks with our immersive visualization with line patterns.
The color blue is for the start of the interaction, and the red is for the end.
While these example patterns for tasks on (a) geo-spatial data distribution, (b) time series, (c) interaction panel, and (d) overall data features are clearly different, it is not clear if the user behaviors are distinguishable for general visual analytics tasks.
However, such observations of different patterns promote our motivation to create classification models of user behaviors for visualization.}
\label{fig:behaviorexamples}
\end{figure*}

\subsection{Visualization Tasks}

Our goal is to study the user behaviors for a diverse set of common visualization tasks and later to interpret unlabeled visual analytics processes with these tasks. 
As shown in Figure~\ref{fig:behaviorexamples}, our study is built on the observations that the user behaviors can be distinguished from some tasks, such as the spatial and temporal tasks; while the differences become subtle to non-distinguishable for detailed tasks inside each visualization space.
Therefore, we strive to cover the most common visualization tasks related to the Covid-19 visualization to study the potential of using ML models to classify general user behaviors during visual analytics.

To ensure the ``uniqueness" of the tasks, we start with the basic visualization task types from an information visualization task taxonomy~\cite{1532136}.
They are all low-level analysis tasks that ``largely capture people’s activities while employing information visualization tools for understanding data"~\cite{1532136}.
These are also the tasks selected for studying eye behaviors related to the visualization domain~\cite{10.1145/3025171.3025187,10.1145/2449396.2449439}.
Different from~\cite{10.1145/3025171.3025187,10.1145/2449396.2449439}, we keep all the task types except ``Filter", as it is a specific interaction type that is not required in our visualization application.
Since all the task types above are with specific goals, we also add a new category ``Exploration" for an interactive exploration process without clear goals, which is very common during visual analytics.
Table~\ref{table:task2} lists the ten categories of our basic visualization tasks, each with a clear description.

To ensure the ``completeness" of the tasks, we organize our tasks based on the involved visualization spaces and tailor all the basic task types to the spatial, temporal and combined spaces.
As shown in Table~\ref{table:task}, the first three task categories are spatial, temporal and interaction -- each corresponds to a visualization space marked `$*$' in the table.
These categories are for tasks that are only needed to be performed with one visualization panel.
The fourth ``combined" space is for tasks that are performed possibly with all the first three visualization spaces.
The indices of each task category combine the numbers of visualization space in parenthesis () and the basic task type from Table~\ref{table:task2}, which are also used in our model methods.

Corresponding to the multi-level task abstraction~\cite{6634168}, we expect that high-level tasks from combined spaces are achieved with lower-level tasks from a single or combined spaces.
Specifically, all the ``combined" tasks may be achieved with a sequence of the ``space", ``time" or ``interaction" tasks.
We expect that users may perform similar basic tasks, but in difference sequences.
Also, our ``combined" tasks are still relatively basic compared to challenging tasks performed during open exploration sessions.
Our goal is still on using the basic user behaviors to interpret the general visual analytics process, with the possible nested complex tasks and open tasks. 

It is worth to mention that while we try to cover a diverse set of basic visualization tasks, our goal is not to come up with an exhaustive list.
Users may always perform some tasks that are unique to their own interests and background.
Instead, we believe that our method can be extended to study user behaviors from additional task types and visualization applications.
We also ignore the challenging levels of tasks, and expect the tasks are completed with different lengths of durations.
Generally, the challenge levels increase from task types 0 to 9.
The ``combined" tasks also require more efforts compared to other single-space tasks.

\section{Comparable Visualization Systems}

We have developed two comparable systems, a desktop and an immersive visualization, for collecting training data in our study.
Both systems support the analysis of Covid-19 dataset with the same set of visualization and interaction functions.
To focus on the general visual analytics process, we did not integrate any advanced algorithms that could accelerate the visualization process, and only provide the basic functions that allow users to visualize any data attribute or select any state at any time stamp. 

\begin{figure}[h]
\centering
\includegraphics[width=1.6in]{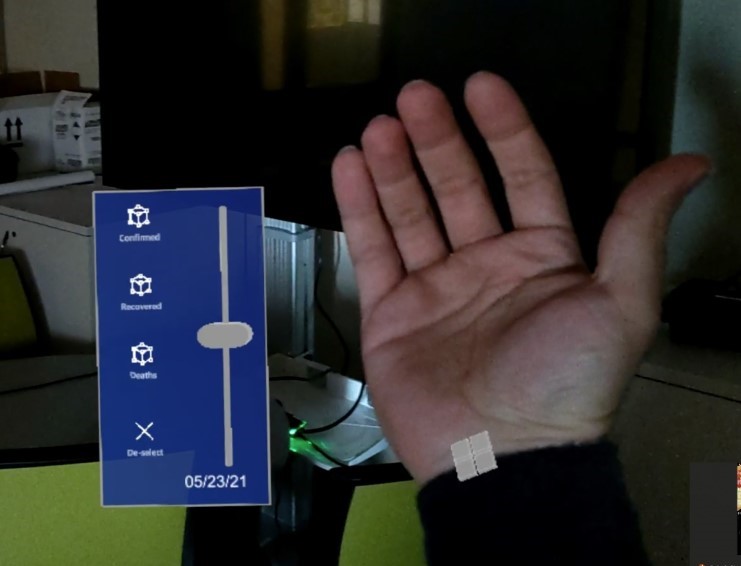}
\includegraphics[width=1.6in]{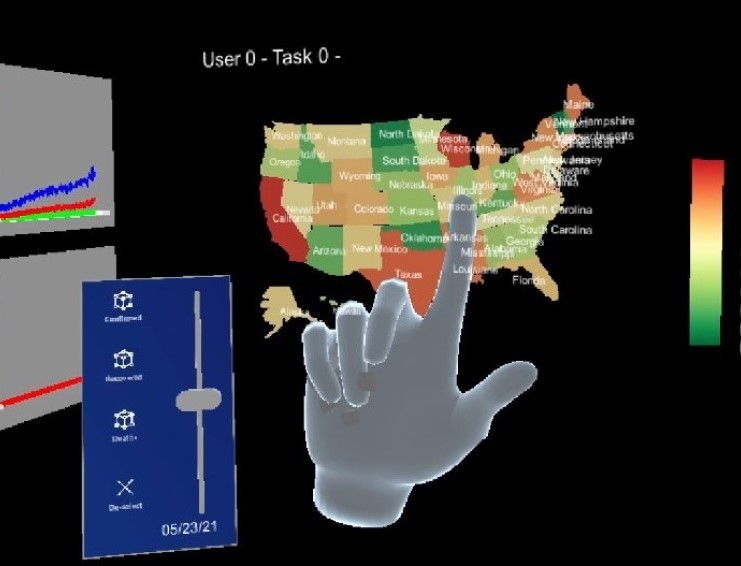}\\
\includegraphics[width=1.6in]{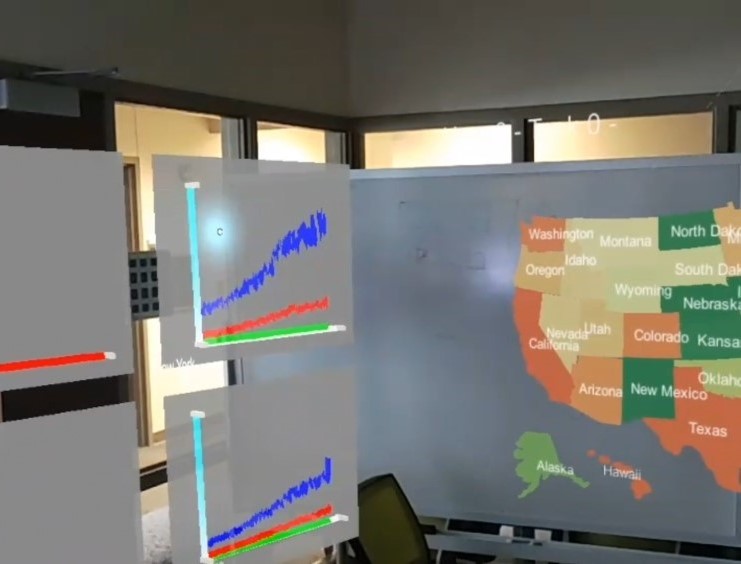}
\includegraphics[width=1.6in]{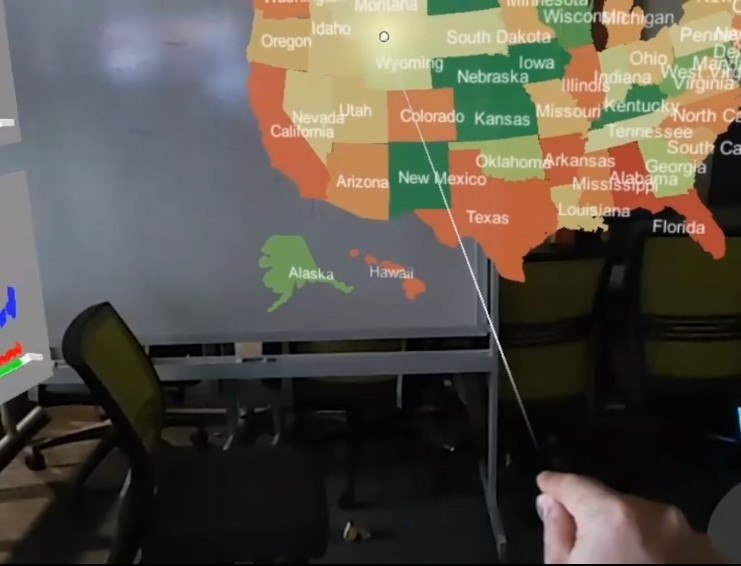}\\
\caption{Interaction examples with the eye tracker and hand gestures in our immersive visualization.
The first row shows two examples of using open palm to view the interaction panel, and the second row shows the eye tracker for participants (lighted dot on the line chart) and a hand gesture `air tap' for selecting a state on map.}
\label{fig:interactions}
\end{figure}

\subsection{Immersive Visualization}
  
The immersive system is developed for Microsoft HoloLens 2 with Unity3D engine.
While both virtual and augmented realities are options of immersive visualization, we have found the most complete data collection from HoloLens 2.
Compared to VR, the AR environment may cause more distraction of users for being able to see the virtual and real worlds simultaneously.
However, none of participants report any issues on distraction or motion sickness.

As shown in Figure~\ref{fig:systems}, both systems contain the following three visualization panels.
\begin{itemize}
\item Choropleth - We adopt a diverging color palette to visualize data values for each state in the US.
The immersive system also includes names of the states in the choropleth.
\item Line charts - The charts visualize the temporal trends of the three attributes for selected states. We support up to 4 charts for side-by-side comparisons.
\item The interaction panel contains buttons for selecting each of the three attributes, a button to remove all selections for both choropleth and time charts, and a slider for adjusting the time stamp for the choropleth.
\end{itemize}

\textbf{Interaction.}
Our system supports the interaction through the combined channels of gaze, gesture and voice.
Specifically, we provide the following interaction functions:

\begin{itemize}
\item Each state can be selected (and de-selected by clicking again) with gazing over a state object and `air tapping' to indicate selection.
\item Each attribute can be selected on the interaction panel and used to update choropleth automatically.
\item The time slider can be selected with `air tap' and moved along the bar to switch time stamp, which automatically updates the choropleth.
\item Each line chart can show the temporal trends for a selected state. The ordering of the four line charts is automatically handled by our system. 
Given a selected state, our system assigns the state to the first available slot.
When the state is de-selected, it is simply removed from the line charts without alternating the other charts.
\item  To support the user study, we provide a function for participates to start or proceed to the next task with a voice command 'next' and view the task description until all tasks are finished.
All the user behaviors and system states during the study are automatically recorded in a log file.
\end{itemize}

\subsection{Desktop Visualization}


Our desktop visualization is built to match the immersive visualization system.
As shown in Figure~\ref{fig:systems}, the desktop system also consists of the three visualization components: the choropleth, line charts and an interaction panel. 

We also maintain the consistency between two systems by processing the data in the same manner and using the same colormap for the choropleth. 
The interaction functions are designed in the exactly same way, except that all are achieved with mouse events.
Specifically, mouse click is used in most interactions, including selecting a state on the choropleth and showing its line chart, clicking a button for switching attributes / clearing all selections, dragging the time slider to a time stamp, or moving to the next trial for the user study.
The names of states are interactively shown or hide when the mouse is placed on top or leave the state. 
The desktop system also logs all user behaviors and system states automatically.


\section{User Study}
We have performed a controlled user study with both visualization systems and we describe the details of the study in this Section.







\subsection{Materials}


The same set of basic visualization tasks described in Section 3.3 was used for both systems.
Among which, the spatial, temporal and combined tasks were exactly the same, while the interaction tasks were dependant on the visualization systems.
Specifically, the immersive system provided 5 interaction functions for selecting an attribute, switching the time stamp, selecting a state, de-selecting a state, and clearing all selections; and the desktop system provided the first 3 interaction functions.

We also designed an open task with the most challenging question -- ``Find 10 insights related to the handling of pandemic and draw a conclusion on how to handle such situations".
The purpose of this task is to collect the testing data to examine our classification models trained from the prior trials.

To balance the effects created by specific task contents, such as the same type of task performed on different state or time stamp, we assigned 3 trials for each task category.
Therefore, the total trial number is 103 for the immersive system and 97 for the desktop version. 

Since combined tasks require the skills from the first three task categories (spatial, temporal and interaction), we placed the combined tasks at the end of the study.
The order of the first three categories should not affect the study results, therefore we follow the indices of task categories in Table~\ref{table:task} for the order of our trials.
As the tasks in each space also follow the order of task types in Table~\ref{table:task2}, the tasks are generally changed from simple to hard in the study. 
To avoid learning effects, our systems randomly pick a dataset for each trial.



\subsection{Participants}

The participants were recruited after an initial screening questionnaire to ensure that they all had normal visions and older than 18.
Among 20 participants (N=20) for the study, 15 were male (75\%) and 5 female (25\%). 
Most of the candidates were college students, with ages ranging from 20 to 45.
Among them, 9 were from the field of Computer Science, 3 from Biology and the rest from other disciplines.

We collected the previous experience of visualization among participants.
In a scale of 1 to 5 (1 being the lowest),
the average score is $3.52$, and
$80\%$ of the participants had scores of  visualization experience from $3-5$.
For the previous experience with VR/AR devices, the average score is $2.52$, and only $45\%$ participants had experience with VR/AR.



\subsection{Procedure}

We first introduced the study with a pre-recorded video, so that all participants were provided with same amount of information about the study.
We also described the procedure of the study, showed the list of tasks and briefly explained their meanings.
Participants were encouraged to ask any questions about the tasks.

Then, the user study proceeded with a practice session for the immersive and desktop systems respectively.
Both practice systems were the same systems used in the study, but with a different dataset.
We instructed participants to practice all the provided interaction functions with the visualization systems, especially the voice and gestures for the immersive visualization.
There was no time limit on the practice session, and participants were asked to get familiarized with the systems until they were ready to begin the user study. No data was collected during the practice trial.

When participants were ready, we instructed them to complete each task and move to the next task immediately.
We also asked participants to speak aloud about what they were doing and their answers to the tasks briefly.
A practical usage of HoloLens is that participants could choose an open space to start the program, and there was no restrictions on making any gestures or moving their bodies while performing the tasks.
We then started voice recording and instructed participants to proceed with the study.
Both our systems led the participants to perform all the trials according to the pre-designed sequence and recorded data automatically.
Participants were not distracted by any data collection during the study.

The orders of immersive and desktop systems were randomized among participants to balance the learning effects on the tasks.
We monitored the study quietly in case any question came up.  
The times of questions were recorded manually so that we could remove the data from these durations.
After the study, we downloaded the log file from the devices.



\subsection{Apparatus}

Our study used Microsoft HoloLens 2 for immersive visualization and a 17 inch laptop for desktop visualization.
The study was performed in a room with an 3x3 meters open space, where participants could move freely during the immersive session. 
For desktop session, participants sat in front of the desktop for the study.
The data for immersive system was captured at the maximum system refresh rates - around 30 frames per second (fps).
The data for desktop was captured at 10 timers per second for mouse locations and real-time for the mouse clicks. 
The voices were recorded with `Jabra speak for conferences' for the immersive system, and the laptop for the desktop system.

\section{Classification of User Behaviors}

In this section, we start with the problem formulation to define our work into a multi-variate time series classification problem. 
Second, we summarize and categorize all the features from the data provenance we gathered in the user study described in Section 5. 
Third, we introduce the transformation techniques that help to extract statistic features from the behavior series and explore the potential options for the machine learning-based methods to tackle this classification problem. 

\subsection{Notations and Problem Formulation}

The prediction of the user's task on visual interface can be essentially viewed as a time-series classification problem \cite{shumway2000time}. 
We define the input data as $n$ series of time-ordered values, $X_i = [{x_i}^{(1)},{x_i}^{(2)},...,{x_i}^{(T_i)}]$, where $T_i$ is the length of time series with index $i$, ${x_i}^{(t)} \in R^{d}$ which is a feature vector generated by concatenating $d$ dimensions of data features in our provenance at each time step.
Our assigned task categories, the category indices in Table~\ref{table:task}, compose the labels $Y = [y_1, y_2,...y_n]$ for each data series.

We then formulate our classification problem as follows.
Given a training set $S = \{(X_1,y_1), (X_2,y_2),...,(X_n,y_n)\}$, composed of $n$ time-series $X_i$ (behavior vectors) and their associated labels $y_i$ (task category), train a classification model to learn a function $h$ such that $h(X_i) = y_i$.
Note that through exploring such ML models, we strive to answer a fundamental problem to visualization -- if there exists distinguishable differences among user behaviors that can be captured with classification models.


\subsection{Data Collection and Feature Selection}


To compare the user behaviors between the immersive and desktop environments, we collect all the attributes that are potentially distinguishable among tasks. 
The desktop system mainly collects user behaviors with the mouse events and interactions with the visualization, and the immersive system collects a variety of user behaviors including physical movements and eye tracking data.

Based on the different collections of attributes, we organize all the behavior features in three groups: interaction, selection and immersive features.
Among which, the first two groups are shared between the desktop and immersive visualizations, and the last group is unique for the immersive visualization.


The interaction features include the user’s very basic interactions with the objects on the interface, such as the click of an attribute button.
We record both the event of the interaction and the input method of the interaction at the exact time stamp.
It is always achieved by mouse in the desktop system, but can be through the channels of voice, gaze, or hand gestures in the immersive system.
Most of the features in this group can be represented with a binary variable. 
It takes 1 when the interaction having been performed on a targeted object at a time stamp and 0 otherwise. 
In addition, we keep track of the mouse movement as a continuous interaction feature for the desktop environment. 

The selection features represent the features that are selected by user with interactions. 
These features reflect the user's attention on the visualizations while they performing visualization tasks. 
Specifically, we collect the name and position of each state being selected on the choropleth, and the names of the states being shown on the line charts.

The immersive features include all the behavior features that are only available through immersive devices.
We are able to track the precise positions of users and objects in the 3D physical 
space. 
Specifically, we have collected the 3D vectors of the user position, 
the forward direction
and the up direction, which are valid throughout the study. 
We also collect the object position, which can be selected by eye tracking or hand gesture.
The value is only valid during the interaction, otherwise reset to $\overrightarrow{0}$).

While we also collect a set of attributes that mark the status of the visualizations, including the dataset used for each trial, we focus on the three groups of behavior features for the classification.
In summary, we use 12 as the feature dimension $d$ for the desktop visualization and 36 for the immersive visualization.

\subsection{Data Transformation}

One major challenge for this work is the transformation of the data for efficiency and feature preserving.
A practical issue is that ML models are often trained in batches by feeding inputs with the same lengths.
However, the length of $X_i$ for each trial from different participant varies significant, since the data series record real user behaviors. 
The wide range of the invariant time series length requires extra processing on the time-series data. Also, the excessive irrelevant features may lead to over-fitting problem. 

To align the time series data, we have considered two options.
The first option is to preserve data granularity and cut long sequences into multiple ones with the assigned length.
The second option is to preserve the overall patterns of behaviors by accumulating data features from long sequences.
Specifically, we divide each sequence into $l$ segments evenly, where $l$ is the assigned length.
All the data in each segment is accumulated and becomes the new feature vector.
We have tested both options and found better results with the second option on preserving the overall patterns.

To capture the features of behaviors, we also enrich our data with statistic-based features~\cite{christ2018time}. 
We extract several basic statistics measurements for each segment of the data, including mean and standard deviation, as they are shown to be effective for similar classification problems~\cite{spiller2021predicting}. 
The transformed data is in consistent length and dimension with the other attributes, so that they can be handle by ML classification models.

\subsection{Data Cleaning and Filtering}

The datasets collected based on human behaviors can be messy and misleading \cite{spiller2021predicting}. 
In our study, undesired behaviors may be included by participants if they did not stick to the assigned tasks.
This may be caused unintentionally, e.g. if participants got confused by the tasks or distracted by others.
Therefore, it is important to clean the training data by removing undesired behaviors.

We construct several filters to clean the data automatically based on our expectations of needed behaviors.
The first filter is the minimum length of each sequence. 
According to our observation, majority participants spent generally 5 to 30 seconds for each task.
Therefore, we safely drop those short series that are too short (two seconds were used as our threshold). 
Second, we assign several `golden' questions to check if the behavior records contain the necessary components that we expect.
For example, for the task ``summarize average value of states in the west coast in the second half of the duration", we expect that at least one state was selected and the time slider would be placed to the second half of the time interval. 
The sequences that do not pass any of the above filters are removed from the training data.


We also normalized all the continuous features within the range of $[0, 1]$ using the min-max method. 
The interaction tasks are not included in the final data, since they are much shorter than other tasks and can be identified directly by the interaction events.
The final shapes 
\color{black}
(the sample size, the sequence length, the feature dimension) 
\color{black}
of our processed datasets are (1198,100,36) for desktop and (1741,100,69) for the immersive environment. 
It is clear that the immersive data collection offers much more data samples than the desktop version with a higher frequency and additional immersive features.

\subsection{Classification Models}

We have surveyed machine learning methods for time-series classification problems~\cite{bagnall2017great,lstm1,lstm2} and selected the following three models that are suitable for this study. 
We also split the training datasets into a training set and a test set using the 5-fold cross-validation process. 

The first model is the k-nearest neighbor (kNN) classifier with the dynamic time warping (DTW) distance function. The kNN has been proved as one of the most popular and traditional time series classification approach \cite{bagnall2017great}. 

The second model is a shallow neural network with two 1D convolutional layers. As the 1D convolutional neural network (CNN) has proved capable of processing the time-series data, this relative simple neural network model can reach high accuracy while maintaining great computation efficiency \cite{kiranyaz20211d}. 

The third model we choose is the state-of-art time classification model:RandOm
Convolutional KErnel Transform (ROCKET) \cite{dempster2020rocket}. The ROCKET extracts features from time-series data based on randomly generated convolutinoal kernels and combine the features with more general classification model such as the Ridge regression classifier. According to \cite{dempster2020rocket}, the ROCKET beats competitive models with much more complicated structure such as the deep neural network and achieves the best performance on the benchmark time-series classification dataset \cite{dau2019ucr}. 

\begin{table}
\centering
Hyperparameters of the model and the experimental setting\\
\begin{tabular}{ |p{4cm}|p{2.5cm}| }
 \hline
Parameter & Value\\
 \hline
 \hline
N$_{k}$ (kNN) & 30\\
 \hline
F$_{N}$ (CNN)& 10\\
\hline
F$_{L}$ (CNN)& 10\\
 \hline
Convolution layers (CNN)& 2\\
\hline
Epoch number (CNN)& 100\\
\hline
N$_{R}$ (ROCKET)& 5000\\
\hline
Batch size& 32\\
\hline
\end{tabular}
\caption{Default parameters of the proposed methods and the experimental setting.}
\label{table:parameter}
\end{table}

Compared to CNN, both kNN and ROCKET actually have very few hyperparameter to be tuned. We choose the 30 as the number of nearest neighbors for kNN (N$_{k}$) and 5000 as the number of random convolutional kernels (N$_{R}$) for ROCKET model. For the CNN model, the determinations of the two parameters, the filter number F$_{N}$ and the filter length F${_{L}}$ are depended on the specific task. Although larger filter size and number generally lead to higher prediction accuracy, the computational burden can increase significantly. Therefore, as indicated in the Table~\ref{table:parameter}, we choose medium values of these two parameters as F$_N$ = 10 and F$_L$ = 10 in the CNN model.

\begin{figure}[h]
\centering
 \includegraphics[width=3.0in]{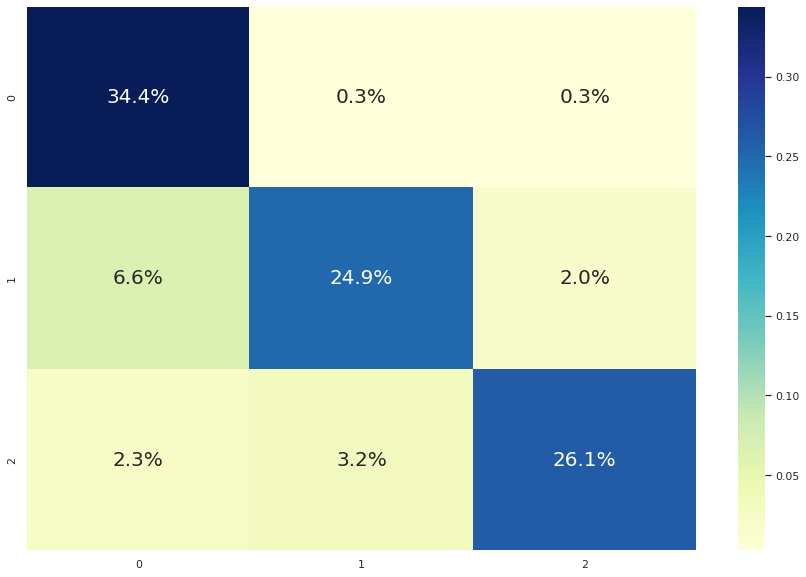}
\caption{Confusion matrix of the kNN model's classification results on the spatial (0), temporal (1) and combined (2) spaces for immersive visualization.
}
\label{fig:Confusion matrix}
\end{figure}

\begin{figure}[h]
\centering
 \includegraphics[width=3.0in]{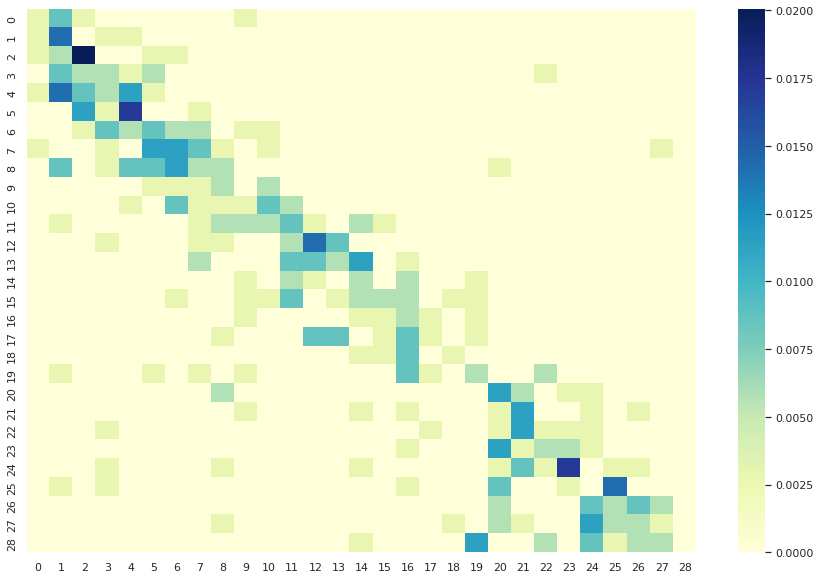}
\caption{Confusion matrix of the kNN model's classification results on the overall 30 task categories from the spatial, temporal and combined spaces for immersive visualization.
}
\label{fig:Confusion matrix_}
\end{figure}

\begin{figure*}[htb]
\centering
 \includegraphics[width=6.0in]{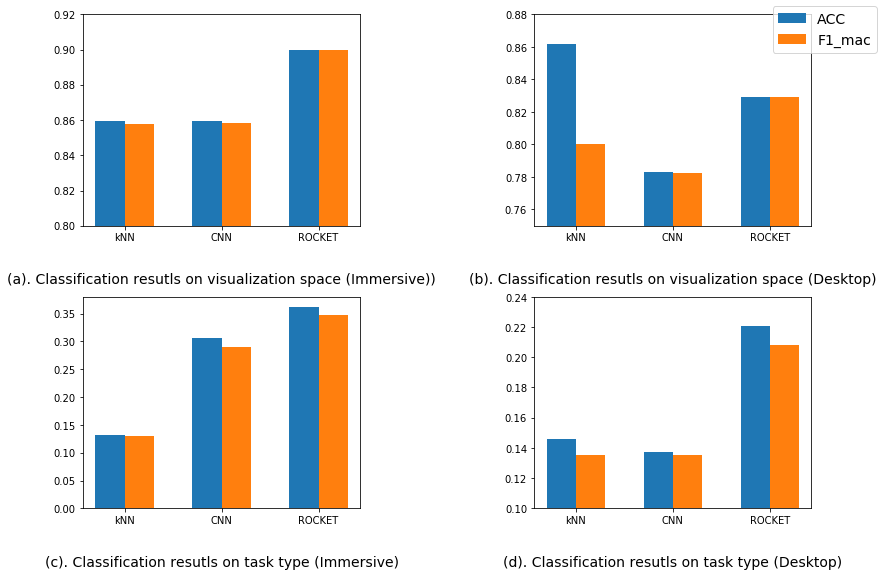}
\caption{The overall performance of kNN, CNN and ROCKET models for immersive (first column) and desktop environment (second column).
The first row shows the classification results for the 3 visual spaces, and the second row shows the results for all the 00-29 tasks in Table~\ref{table:task}.
}
\label{fig:classification}
\end{figure*}

\section{Data Analysis and Results}

\subsection{Classification Results}


We study the classification problems at 2 scales for the 2 environments, and we summarize the results based on these 4 cases.
The `task' scale is for identifying task categories, and the `space' scale is only for classifying the visual space which combines all task types performed for that space.
Both the visual space and category no. are listed in Table~\ref{table:task}.
To evaluate the performance of classification models, we use two classification metrics: the accuracy and the F1$_{mac}$ score. 

Figure~\ref{fig:Confusion matrix} shows the visual space classification results of kNN for the immersive environment.
In this task, the kNN reaches 85.4\% accuracy and F1$_{mac}$ score is 85.2\%. 
Figure~\ref{fig:Confusion matrix_} shows the task classification results of kNN for the immersive environment. At this detailed scale, the accuracy and F1$_{mac}$ scores have dropped to 14.6\% and 13.5\% correspondingly.
It is noticeable that the kNN classifier has difficulty to distinguish the class types within each visual space (0~9, 10~19 and 20~29).
However, the matrix still shows a good classification between different visual spaces, and there are clear separations of detailed task categories. 

A detailed comparisons for all the models' classification results are presented in Figure~\ref{fig:classification}. 
Several facts can be observed here. 
First, with more samples and features collected from the immersive environment, all three models achieve better results than the desktop version. Also, among all three models, the ROCKET outperforms the other two significantly. 
For the visual space classification from immersive visualization, it achieves average accuracy score as 90.0\% and the F1$_{mac}$ score is 90.0\% .
For the task type classification problem, it achieves an average accuracy score of 36.1\% and the F1$_{mac}$ score is 34.8\%. 
The kNN and CNN's performances are very close in most cases. 
The only notable difference between these two models' predictions is during the task type classification for immersive environment. 
In this case, the CNN beats the kNN with 30.7\% average accuracy and 30.0\% average F1$_{mac}$ score, although still way behind the ROCKET's performance. 

\begin{figure*}[htb]
\centering
 \includegraphics[width=6.0in]{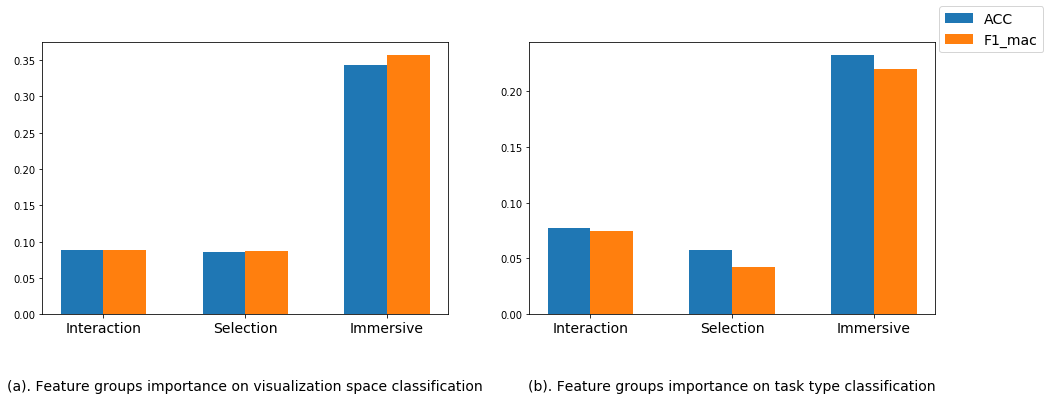}
\caption{
The feature group importance (the drop of the ACC and F1$_{mac}$ scores of ROCKET prediction with leaving the group out) for the space and task classification problems.
}
\label{fig:permutation}
\end{figure*}

\subsection{Importance of Behavior Attributes}

We also analyze the importance of individual attributes for the classification results.
In the other words, what behaviors make the differences during the process of visual analytics?
Since the ROCKET classifier achieves the best accuracy, we focus on this model and its prediction results.
We also perform the analysis with the immersive data collection, since it contains all three groups of behavior features. 

We start with identifying the most important feature group that help the ROCKET model to make the correct predictions. 
To keep this process simple, we leave one group of features out and use the remained features to train the model. 
As shown in Figure~\ref{fig:permutation}, 
the absent of the immersive features let a huge model performance gap for both the space and task classification problems. 
This result is consistent with the findings shown in Figure~\ref{fig:classification}, which reveals that the immersive features recorded in the immersive environment are crucial for improving the models' performance.

\begin{table}
\centering
\begin{tabular}{ |p{2cm}|p{2cm}| }
 \hline
Position & Feature \\
 \hline
 \hline
1 & objPosition.y \\
 \hline
2 & position.x \\
\hline
3 & objPosition.z \\
 \hline
4 & up.x \\
\hline
5 & forward.z \\
\hline
6 & up.z\\
\hline
7 & up.y\\
\hline
8 & position.y\\
 \hline
9 & position.z\\
\hline
10 & forward.y\\
 \hline
11 & forward.x\\
\hline
12 & objPosition.x\\
\hline

\end{tabular}
\caption{The importance ranking of immersive features based on PermFIT.}
\label{table:feature importance}
\end{table}

We then zoom on to the 12 features in the immersive group. 
They are the world position of the object participants interacted with (objPosition.x, objPosition.y, objPosition.z), the world position of the HoloLens HMD (position.x, position.y, position.z), the front direction of the participant (forward.x, forward.y, forward.z) and the UP direction of the participant (up.x, up.y, up.z) which measures the head orientation. 
To further examine the influence of each immersive feature, we implement the feature permutation importance technique on the ROCKET model for the space classification problem~\cite{good2013permutation, mi2021permutation}. 
The Permutation-based Feature Importance Test (PermFIT) algorithm~\cite{mi2021permutation} is used to train the ROCKET model again. 
Each training process is conducted via the 5-fold cross-validation while permuting one of the immersive feature in the dataset. 
We repeat the process for 100 times and check the the expected squared difference between the outcomes with the original input $X_{i}$ and the permuted one $X_{i}^{'}$. 
This difference is considered as the importance score for the corresponding feature. 
Table~\ref{table:feature importance} summarizes our results. 
It reveals that the object positions (objPosition.y, objPostion.z), such as the coordinate of the selected state in the choropleth, have significant impact on the model performance. Also, the participants' head orientation (up.x, up.y, up.z) gives great information about what they are actually doing during the visualization process. 

\subsection{Interpretation of Exploration Process}

\begin{figure*}[htb]
\centering
 \includegraphics[width=6.0in]{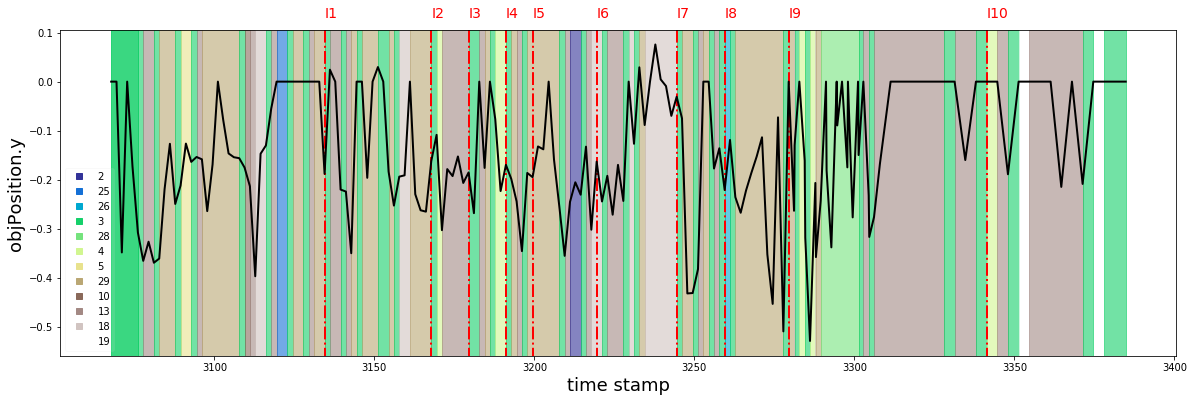}
\caption{An example of automatic interpretation of the visualization process. The colors mark the task categories identified with our model. The time line shows the associations of a physical behavior feature (the black line) and the voice transcript (the 10 insights reported by the participant).
}
\label{fig:exploration}
\end{figure*}

We demonstrate one usage of our classification models by using them to interpret the exploration process of the open problem in the last trial of our study.
To reach this purpose, we use the well-trained ROCKET model at the detailed task scale. 
Figure~\ref{fig:exploration} visualizes a specific exploration trace for one participant (participant 7) with the immersive environment. 
Based on our analysis results, we choose the most significant feature, the objPosition.y, as the black indicator in the figure.
To generate the testing instances, we use a growing window, which starts with containing the first 100 timestamps and gradually reaching the full sequence length, to segment the whole sequence. The regenerated sequences are processed in the same manner as the training data.
The classification results reveal that this visualization process mixes a number of visualization tasks from all the three visual spaces and 12 task categories can be identified with high confidence.
By transcribing the voice record, we recovered 
these tasks including finding extreme value, sorting dataset, and characterizing data distribution.
This example demonstrates the effects of ML models for automatic interpretation of visualization process.


\section{Discussions and Conclusion}

\subsection{Discussions}

The purpose of our study is to study the problems of user behaviors during the processes of visual analytics, instead of comparing the effectiveness or efficiency between immersive and desktop systems. 
Our focus is on studying if user behaviors are different and what behaviors make the differences, and expect that each environment would be able to distinguish a subset of user behaviors.

As multi-attribute machine learning methods have been proven to be more effective in many cases~\cite{10.1145/3301400, 10.1145/3446638}, immersive systems provide an excellent study environment to capture various detailed user behaviors.
The differences of data collections are obvious from the two environments, and the classification results are as expected.
Our data collection contains all the commonly used data of behaviors, including the eye tracking data, which increases the chances of identifying effective behavior factors.

Our study with immersive visualization provides a 3D layout in the physical space, which encourages participants' physical movements such as walking toward the choropleth or line charts to observe the details.
The physical movements are scattered in the entire 3D physical space, which boosts the performance of automatic algorithms.
While physical behaviors are used to be ignored in the previous studies, we demonstrate their functions on understanding user behaviors for visualization tasks.

The data collected with our study can be used for additional problems and our models can be further improved by refining the behavior patterns.
The discrepancy on the understanding on the tasks always exist among participants, especially for people with different background.
Even for the basic geospatial tasks performed in the study, participants with and without the visualization background perform differently.
With approaches to studying individual differences, we expect the results can produce personalized models and overall better performances for a group of users as well.

\subsection{Conclusion and Future Work}
This work presents a comparison study on user behaviors when they perform visual analytics tasks between desktop and immersive visualizations.
Our work collects a comprehensive set of interaction and visualization system provenance at a very high frequency with the latest AR devices.
We use a classical visualization problem -- geospatial, temporal and multi-attribute visualization -- to study how users perform common visual analytics tasks. 
We have explored the latest time-series classification models to capture important features that differentiate user behaviors during different visualization tasks, and we illustrate the usage of our models by automatic interpreting the open exploration processes of users.
Our results demonstrate that time-series classification models provide useful information to evaluate existed visualization systems and design further intelligent interactive analytics functions.
Since our classification models can adapt to various feature sets and multivariate attributes from sequences of user behaviors, we plan to expand our study to more tasks of immersive and visual analytics.

In the future, we plan to extend this work from three directions. 
First, we will explore additional ML models that can distinguish more subtle differences among the sequences of user behaviors for visualization problems. 
Second, we will continue to study other usages of such ML models on problems, such as studying the complex sensemaking procedures of interactive analytics with the support of automatic and robust annotation.
Third, we are interested in training ML models with small datasets, as the data collection is often a time consuming process even with automatic annotations.
While this work studies geo-spatial visualization, we believe that the method of our study can be extended to various visualization task types.

\section{Acknowledgement}
This work was supported by NSF Award 1840080. 
The authors thank for the discussions of related ideas with Dr. Jian Huang and Dr. Xintao Wu. 
Shahin Doroudian contributed to the user study with the HoloLens and participated in the early discussions of this project.

\bibliographystyle{abbrv-doi}

\bibliography{template}
\end{document}